\documentclass[preprint]{elsarticle}

\usepackage{graphicx} 
\usepackage[loose]{subfigure}      
\usepackage{dcolumn}  
\usepackage{bm}       

\newcommand*{\meg}            {\mu \to e \gamma}
\newcommand*{\megsign}        {\mu^+ \to e^+ \gamma}

\newcommand*{\egamma}         {E_{\gamma}}
\newcommand*{\epositron}      {E_{\rm e}}
\newcommand*{\tegamma}        {t_{{\rm e}\gamma}}

\newcommand*{\dtheta}         {\theta_{e \gamma}}
\newcommand*{\dphi}           {\phi_{e \gamma}}
\newcommand*{\nsig}           {N_{\rm sig}}
\newcommand*{\nenunu}         {N_{e \nu \bar \nu}}
\newcommand*{\nrd}            {N_{\rm RMD}}
\newcommand*{\nbg}            {N_{\rm BG}}
\newcommand*{\nobs}           {N_{\rm obs}}

\begin{document}


\title{A limit for the {$\mu \to e \gamma$} decay from the MEG experiment}

\author[]{MEG Collaboration}

\author[PSI,ETH]   {J.~Adam}
\author[ICEPP]     {X.~Bai}
\author[INFNPi]    {A.~Baldini$^{\S}$}
\author[UCI]       {E.~Baracchini}
\author[INFNRm]    {A.~Barchiesi}
\author[INFNPi]    {C.~Bemporad$^{\S\dagger}$}
\author[INFNPv]    {G.~Boca$^{\S\dagger}$}
\author[INFNPv]    {P.~W.~Cattaneo$^{\S}$}
\author[INFNRm]    {G.~Cavoto$^{\S}$}
\author[INFNPv]    {G.~Cecchet$^{\S}$}
\author[INFNPi]    {F.~Cei$^{\S\dagger}$}
\author[INFNPi]    {C.~Cerri$^{\S}$\fnref{retired}}
\author[INFNPv]    {A.~De~Bari$^{\S}$}
\author[INFNGe]    {M.~De~Gerone$^{\S\dagger}$}
\author[Waseda]    {T.~Doke}
\author[INFNGe]    {S.~Dussoni$^{\S\dagger}$}
\author[PSI]       {J.~Egger\fnref{retired}} \fntext[retired]{retired}
\author[INFNPi,PSI]{L.~Galli$^{\S\dagger}$}
\author[INFNPi,PSI]{G.~Gallucci$^{\S\dagger}$}
\author[INFNGe]    {F.~Gatti$^{\S\dagger}$}
\author[UCI]       {B.~Golden}
\author[INFNPi]    {M.~Grassi$^{\S}$}
\author[BINP]      {D.~N.~Grigoriev}
\author[KEK]       {T.~Haruyama}
\author[PSI]       {M.~Hildebrandt}
\author[ICEPP,PSI] {Y.~Hisamatsu}
\author[BINP]      {F.~Ignatov}
\author[ICEPP]     {T.~Iwamoto}
\author[ICEPP]     {D.~Kaneko}
\author[PSI]       {P.-R.~Kettle}
\author[BINP]      {B.~I.~Khazin}
\author[PSI]       {O.~Kiselev}
\author[JINR]      {A.~Korenchenko}
\author[JINR]      {N.~Kravchuk}
\author[KEK]       {A.~Maki}
\author[KEK]       {S.~Mihara}
\author[UCI]       {W.~Molzon}
\author[ICEPP]     {T.~Mori}
\author[JINR]      {D.~Mzavia}
\author[ICEPP,PSI] {H.~Natori}
\author[INFNPv]    {R.~Nard\`o$^{\S\dagger}$}
\author[INFNPi]    {D.~Nicol\`o$^{\S\dagger}$}
\author[KEK]       {H.~Nishiguchi}
\author[ICEPP]     {Y.~Nishimura}
\author[ICEPP]     {W.~Ootani}
\author[INFNLe]    {M.~Panareo$^{\S\dagger}$}
\author[INFNPi]    {A.~Papa$^{\S\dagger}$}
\author[INFNPi]    {R.~Pazzi$^{\S\dagger}$\fnref{deceased}}        \fntext[deceased]{deceased}
\author[INFNRm]    {G.~Piredda$^{\S}$}
\author[BINP]      {A.~Popov}
\author[INFNRm]    {F.~Renga$^{\S\dagger}$}
\author[PSI]       {S.~Ritt}
\author[INFNPv]    {M.~Rossella$^{\S}$}
\author[ICEPP]     {R.~Sawada}
\author[PSI,ETH]   {M.~Schneebeli\fnref{mattias}}   \fntext[mattias]{Present address: DECTRIS Ltd., Neuenhoferstrasse 107, CH-5400 Baden, Switzerland}
\author[INFNPi]    {F.~Sergiampietri$^{\S}$}
\author[INFNPi]    {G.~Signorelli$^{\S}$}
\author[Waseda]    {S.~Suzuki}
\author[UCI]       {C.~Topchyan}
\author[UCI]       {V.~Tumakov}
\author[ICEPP,PSI] {Y.~Uchiyama}
\author[INFNGe]    {R.~Valle$^{\S\dagger}$\fnref{fnvalle}}        \fntext[fnvalle]{Present address:  Simco S.r.l., 16042 Carasco, Italy}
\author[INFNRm]    {C.~Voena$^{\S}$}
\author[UCI,PSI]   {F.~Xiao}
\author[KEK]       {S.~Yamada}
\author[KEK]       {A.~Yamamoto}
\author[ICEPP]     {S.~Yamashita}
\author[BINP]      {Yu.~V.~Yudin}
\author[INFNRm]    {D.~Zanello$^{\S}$}

\address[INFNPi]{{INFN Sezione di Pisa$^{\S}$; Dipartimento di Fisica$^{\dagger}$ dell'Universit\`a, Largo B.~Pontecorvo~3, 56127 Pisa, Italy}}
\address[INFNGe]{{INFN Sezione di Genova$^{\S}$; Dipartimento di Fisica$^{\dagger}$ dell'Universit\`a, Via Dodecaneso 33, 16146 Genova, Italy}}
\address[INFNPv]{{INFN Sezione di Pavia$^{\S}$; Dipartimento di Fisica$^{\dagger}$ dell'Universit\`a, Via Bassi 6, 27100 Pavia, Italy}}
\address[INFNRm]{{INFN Sezione di Roma$^{\S}$; Dipartimento di Fisica$^{\dagger}$ dell'Universit\`a ``Sapienza'', P.le A.~Moro~2, 00185 Roma, Italy}}
\address[INFNLe]{{INFN Sezione di Lecce$^{\S}$; Dipartimento di Fisica$^{\dagger}$ dell'Universit\`a, Via per Arnesano, 73100 Lecce, Italy}}
\address[ICEPP]{{ICEPP, The University of Tokyo 7-3-1 Hongo, Bunkyo-ku, Tokyo 113-0033, Japan }}
\address[UCI]  {{University of California, Irvine, CA 92697, USA}}
\address[KEK]   {{KEK, High Energy Accelerator Research Organization 1-1 Oho, Tsukuba, Ibaraki 305-0801, Japan}}
\address[PSI]   {{Paul Scherrer Institute PSI, CH-5232 Villigen, Switzerland}}
\address[ETH]   {{Swiss Federal Institute of Technology ETH, CH-8093 Zuerich, Switzerland}}
\address[Waseda]{{Research Institute for Science and Engineering, Waseda~University, 3-4-1 Okubo, Shinjuku-ku, Tokyo 169-8555, Japan}}
\address[BINP]   {{Budker Institute of Nuclear Physics, 630090 Novosibirsk, Russia}}
\address[JINR]   {{Joint Institute for Nuclear Research, 141980, Dubna, Russia}}


\date{\today}

\begin{abstract}
A search for the decay $\mu^+ \to e^+ \gamma$, performed at PSI and
based on data from the initial three months of operation of the MEG
experiment, yields an upper limit on the branching ratio of
BR$({\mu^+ \to e^+ \gamma}) \leq 2.8 \times 10^{-11}$ (90\% C.L.).
This corresponds to the measurement of positrons and photons from
$\sim 10^{14}$ stopped $\mu^+$-decays by means of a superconducting
positron spectrometer and a 900$\,$litre liquid xenon photon
detector.
\end{abstract}

\maketitle
\section{Introduction}
We report here on the results of a search for the lepton flavour
violating decay $\mu^+ \to e^+ \gamma$, based on data collected
during the first three months period of the MEG experiment. This
operates at the 590$\,$MeV proton ring cyclotron facility of the
Paul Scherrer Institut (PSI), in Switzerland.

Lepton flavour conservation in the Standard Model $\,$(SM) is
associated with neutrinos being massless. Recent observations of
neutrino oscillations~\cite{nuosc} imply a non-zero mass and hence
the mixing of lepton flavours. However, in minimal extensions to the
SM, with finite but tiny masses, charged lepton flavour violating
processes are strongly suppressed and beyond experimental reach.

Additional sources of lepton flavour violation (LFV)~\cite{Barbieri,
hisano,LFreview} appear in theories of supersymmetry, grand
unification or in extra dimensions, giving predictions that have now
become accessible experimentally. Hence, the present lack of
observation of a signature of charged LFV may change with improved
searches and reveal new physics beyond the SM or significantly
constrain the parameter space of such extensions.

The strongest bounds on charged LFV come from the muon system, with
the current limit for the branching ratio BR$(\megsign) \leq 1.2
\times 10^{-11}$ (90$\%$~C.L.), set by the MEGA
experiment~\cite{MEGA}.

\section{Experimental Principle}
The $\megsign$ process is characterized by a simple two-body final
state, with the positron and photon being coincident in time and
emitted back-to-back in the rest frame of the muon, each with an
energy equal to half that of the muon mass.

There are two major sources of background, one from radiative muon
decay (RMD) $\mu^+ \to e^+ \nu_e \bar\nu_\mu\gamma$ and the other
from accidental coincidences between a high energy positron from
the normal muon decay $\mu^+ \to e^+ \nu_e \bar\nu_\mu$
(Michel decay) and a high energy photon from sources such as RMD,
positron annihilation-in-flight or bremsstrahlung. Both types of
background can mimic a $\meg$ event by having an almost back-to-back
photon and positron. It can be shown~\cite{kuno-okada}, taking into
account the muon rate as well as the acceptances and resolutions,
that the accidental case dominates.

Hence the key to suppressing such backgrounds lies in having a
continuous muon beam, a good quality beam transport system and
precision detectors with excellent spatial, temporal and energy
resolutions. This is the basis for the novel design of MEG.
\section{Experimental Layout and the MEG Detector}
\begin{figure*}[ht]
\begin{center}
\includegraphics [width=0.98\linewidth]{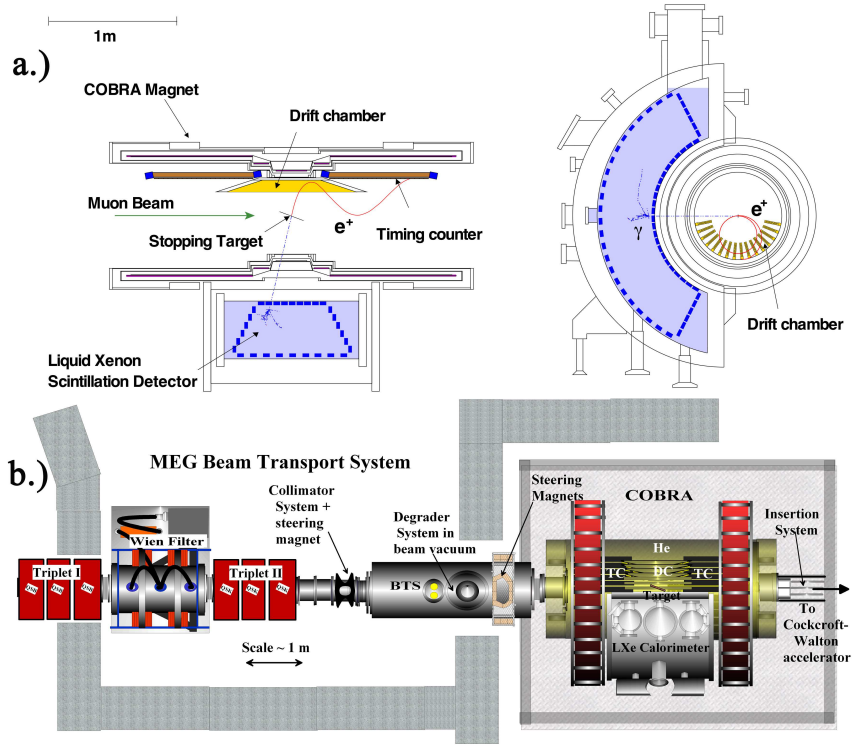}
\end{center}
\caption{\label{fig:sketch}Schematic of the MEG experiment. a)
details of the COBRA positron spectrometer and LXe detector, showing
the positron tracking chambers (DC), the scintillator timing counter
arrays (TC) and the superconducting gradient-field solenoid COBRA.
b) The MEG beam transport system, surface muons enter from the left.
Also shown are the crossed-field separator (Wien filter), the
superconducting transport solenoid (BTS) and the COBRA magnet with
the central thin stopping target surrounded by the various
detectors. }
\end{figure*}

A schematic of the experiment is shown in Figure~\ref{fig:sketch}.
Surface muons of 28$\,$MeV/c from one of the world's most intense
sources, the $\pi$E5 channel at PSI, are stopped in a thin,
partially depolarizing polyethylene target, placed at the centre of
the positron spectrometer. To facilitate a stopping rate of $3\times
10^7 \mu^+ s^{-1}$ in the 18$\,$mg/cm$^2$ thick target, with minimum
beam-related background, a Wien filter and a superconducting
transport solenoid (BTS) with a central degrader system are
employed. The MEG beam transport system so, cleanly separates
(7.5$\sigma$) the eight times higher positron contamination to
provide a pure muon beam. The use of a helium environment in the
spectrometer, together with a slanted target, ensures  minimal
multiple scattering for both the muons and the out-going positrons.
This is essential also for limiting background production, such as
annihilation-in-flight and bremsstrahlung, in the acceptance region,
centred around 90$^\circ$ to the incoming beam.

Positrons originating from muon decay are analyzed in the COBRA
(COnstant-Bending-RAdius) spectrometer consisting of a thin-walled
superconducting magnet with a gradient magnetic field, a tracking
system of low-mass drift chambers and two fast scintillator
timing-counter arrays.

The gradient magnetic field in the spectrometer, ranging from 1.27
Tesla at the centre to 0.49 Tesla at either end, is designed such
that positrons emitted from the target with the same momentum follow
trajectories with an almost constant projected bending radius,
independent of their emission angle. This allows a preferential
acceptance of higher momentum particles in the drift chambers as
well as sweeping away particles more efficiently, compared to a
uniform field.

The drift chamber system (DCH) consists of 16 radially aligned
modules, spaced at $10.5^\circ$ intervals, forming a half-circle
around the target. Each drift chamber module contains two staggered
layers of anode wire planes each of nine drift cells. The two layers
are separated and also enclosed by 12.5\,$\mu$m thick cathode foils
with a Vernier pattern structure. The chambers are operated with a
helium:ethane (50:50) gas mixture, allowing this low-mass
construction to total $2.0\times10^{-3}$\,$X_0$ along the positron
trajectory.

Positron timing information originates from fast, scintillator
timing-counter arrays (TC), placed at each end of the spectrometer.
Each array consists of 15 BC404 plastic scintillator bars, with 128
orthogonally placed BCF-20 scintillating fibres. Each bar is
read-out at either end by a fine-mesh photomultiplier tube, while
the fibres are viewed by avalanche photo-diodes. The precise timing
and charge signals provide both the impact point on the TCs and give
directional information on the positron.

The photon detector is a 900~litre homogeneous volume of liquid xenon
(LXe) that subtends a solid-angle acceptance of $\sim 10\,\%$. It uses
scintillation light to measure the total energy released by the
$\gamma$-ray as well as the position and time of its first
interaction. In total, 846 photomultiplier tubes (PMTs), internally
mounted on all surfaces and submerged in the xenon, are used.
The advantage of using liquid xenon is its fast response,
large light yield and short radiation length.
Stringent control of contaminants is necessary since the vacuum
ultra-violet (VUV) scintillation light is easily absorbed by water
and oxygen even at sub-ppm levels. The xenon is therefore circulated in
liquid phase through a series of purification cartridges, and in gas
phase through a heated getter. Both the optical properties of xenon
as well as the PMT gains and quantum efficiencies are constantly
monitored by means of LEDs and pointlike $^{\rm 241}$Am $\alpha$-sources
deposited on thin wires stretched inside the active volume. The
detector is maintained at 165~K by means of a pulse-tube
refrigerator with a cooling power of 200~W.

To select matched photon and positron candidates in a high rate,
continuous beam environment and store sufficient information for
offline analysis requires a well matched system of front-end
electronics, trigger processors and data acquisition (DAQ) software.

The front-end electronics signals (2748) are actively split and go
to both the trigger and the in-house designed waveform digitizer
boards. The latter are based on the multi-GHz domino ring sampler
chip (DRS), which can sample ten analogue input channels into 1024
sampling cells each at speeds of up to 4.5$\,$GHz. The sampling
speed for the drift chamber anode and cathode signals is 500$\,$MHz,
while that of the PMT signals from the photon detector and timing
counters is 1.6$\,$GHz. This strategy gives maximum flexibility,
allowing various read-out schemes, such as zero suppression, on an
event-by-event basis for various trigger types. The system achieves
an excellent pile-up recognition, together with  superior timing and
amplitude resolutions, compared to conventional schemes.

The trigger is based on fast information from the two detectors
using PMTs: the liquid xenon photon detector and the positron timing
counters. It makes use of a subset of the kinematic observables from
$\mu$-decay at rest, requiring an energy deposit in the photon
detector in an interval around $52.8\,$MeV, a time coincident
positron hit on the timing counters and a rough collinearity of the
two particles, based on their hit topology. The decay kinematics is
reconstructed by electronics boards arranged in a triple layer
tree-structure. The signal digitization is executed by means of a
$100$~MHz, $10$-bit flash analogue-to-digital converter. A
pre-scaled, multi-trigger event scheme is used for data-taking
allowing calibration, background and signal events to be read-out
together. The typical signal event rate was $5$~Hz, and the total
DAQ rate was $6.5$~Hz, with an average livetime of $84\%$.

In total, nine front-end computers are used for the DAQ, each
sending an event fragment to a central event building computer over a
Gigabit Ethernet link. An integrated slow-control system enables
both equipment control and monitoring.

A detailed GEANT 3.21 based Monte Carlo simulation of the full apparatus
(transport system and detector) was developed and used throughout the
experiment, from the design and optimization of all sub-systems to the calculation of
acceptances and efficiencies.

\section{Monitoring and Calibrations}
The long term stability of the MEG experiment is an essential
ingredient in obtaining high quality data over extended measurement
periods. Continuous monitoring and frequent calibrations are a
prerequisite. Apart from such items as the liquid xenon temperature
and pressure, the drift chambers gas composition and pressure and
the electronics temperature, a number of additional measurements
must be performed to keep the subdetectors calibrated and
synchronized. The three most important are nuclear reactions from a
Cockcroft-Walton (CW) accelerator, radiative muon decay (RMD) runs
and pion charge-exchange (CEX) reaction runs.

Three times a week during normal data-taking, $\gamma$-rays of
moderate energy coming from nuclear reactions of protons on a
Li$_2$B$_4$O$_7$ target are used. Protons of variable energy ($400 <
T_p < 1000$~keV) are produced by a dedicated CW-accelerator placed
downstream of the experiment. The muon stopping target is
automatically replaced by a remotely extendable beam pipe which
places the nuclear target at the centre of the detector. Photons of
$E_\gamma = 17.67$~MeV from $^{\rm 7}$Li($p,\gamma$)$^{\rm 8}$Be
allow the monitoring of the LXe detector energy scale, while
coincident $\gamma$'s from $^{\rm 11}$B($p,\gamma$)$^{\rm 12}$C
($E_\gamma = 4.4, 11.6$~MeV) detected simultaneously by the timing
counter and the xenon detector allow the determination of time
offsets of TC bars.

Once a week an entire day of RMD data-taking at reduced beam
intensity was performed, with the trigger requirements
relaxed to include non back-to-back positron-photon events in a
wider energy range.

Two CEX runs ($\pi^- p \to \pi^0 n \to \gamma \gamma n$) were also
conducted, one at the beginning and one at the end of the data-%
taking period. Pion capture at rest in a liquid hydrogen target
produces photons with energy $54.9 < E_\gamma < 83.0$~MeV. By
detecting one of these photons with the LXe detector and the other
at 180$^\circ$ by means of a set of NaI crystals, preceded by a
lead/scintillator sandwich, two mono-energetic calibration lines at
the extremes of the energy spectrum are obtained. These enable
measurement of the energy scale and uniformity. Dalitz decays
($\pi^0 \to \gamma e^+ e^-$) were also collected by using a
photon-positron coincidence trigger, and used to study the detector
time synchronization and resolution.

The combined use of all these methods enables the investigation of
possible systematic variations of the apparatus.

\section{Event selection and resolutions}
The data sample analyzed here was collected between September and
December 2008 and corresponds to $\sim 9.5 \times 10^{13}$ muons
stopping in the target. At the first stage of the data processing, a
data reduction is performed by selecting events with conservative
criteria that require the time of the photon detector signal to be
close to that of a timing counter hit, and at least one track to be
detected by the drift chamber system. This reduces the data size to
16\,\% of the recorded events. The pre-selected data are again
processed and those events falling into a pre-defined window
(blinding-box), containing the signal region on the $\gamma$-ray
energy and the time difference between the $\gamma$-ray and the
positron, are ``hidden'', {\it i.e.} written to a separate
data-stream, in order to prevent any bias in the analysis procedure.
Only the events outside the blinding-box are used for optimizing the
analysis parameters and for studying the background.

During the course of the data-taking period the light yield of the
photon detector was continuously increasing due to the purification
of the liquid xenon, which was performed in parallel. Furthermore,
an increasing number of drift chambers suffered from frequent
high-voltage trips resulting in a reduction of the positron
detection efficiency by a factor of three over the period. The
increase of the xenon light yield was carefully monitored with the
various calibration tools and it is taken into account in the
determination of the energy scale. The trigger thresholds were also
accordingly adjusted to guarantee a uniform efficiency. For the
drift chamber system  we adopted a normalization scheme which
depends only on the ratio of the signal positron reconstruction
efficiency relative to that of the Michel positron, in order to be
insensitive to absolute efficiencies.

A candidate $\mu^+ \to e^+ \gamma$ event is characterized by the
measurement of five kinematic parameters: positron energy ($E_e$),
photon energy ($E_\gamma$), relative time between the positron and
photon ($t_{e \gamma}$) and opening angles between the two particles
($\theta_{e \gamma}$ and $\phi_{e \gamma}$).

\subsection{Positron energy, $E_{e}$}
The positron track is reconstructed  with the Kalman filter technique
\cite{Kalman}, in order to take into account the effect of the multiple
scattering and energy loss in the detector materials in the variable magnetic field.

\begin{figure}
\begin{center}
\includegraphics[width=7cm]{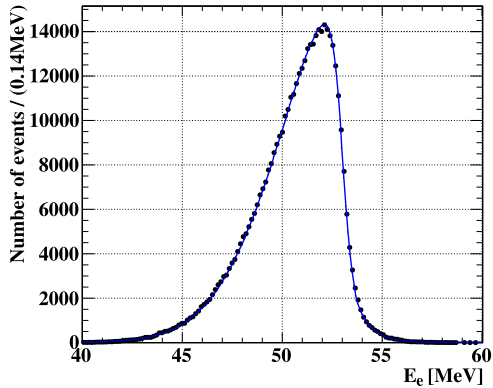}
\end{center}
\caption{\label{fig:Michel spectrum} Measured Michel positron energy
spectrum. A solid line shows the fitted function as described in the
text.
}
\end{figure}

The positron energy scale and resolution are evaluated by fitting
the kinematic edge of the measured Michel positron energy spectrum
at 52.8\,MeV as shown in Figure~\ref{fig:Michel spectrum}. The fit function is formed by folding the theoretical
Michel spectrum form with the  energy-dependent detector efficiency,
and the response function for mono-energetic positrons. The latter
is extracted from the Monte Carlo simulation  of $\mu \to e \gamma$ decays,
and is well described by a triple Gaussian function (a sum of a core and
two tail components).

The resolutions extracted from the data are $374$~keV, $1.06$~MeV
and $2.00$~MeV in sigma for the core component and the two tails,
with corresponding fractions of $60$\,\%, $33$\,\% and $7$\,\%,
respectively. The uncertainty on these numbers is dominated by
systematic effects and was determined by varying both the event
selection and fitting criteria.

\subsection{Photon energy, $E_\gamma$}

\begin{figure}
\begin{center}
\includegraphics[width=8cm]{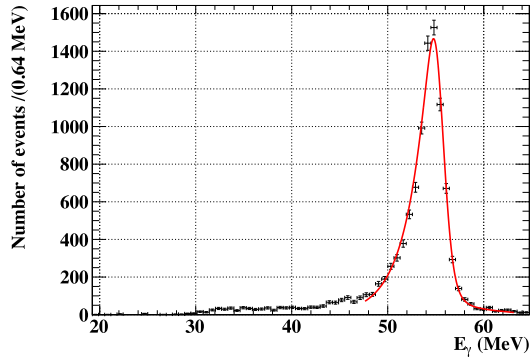}
\end{center}
\caption{\label{fig:CEX spectrum} Measured energy spectrum for
$54.9$~MeV photons from a CEX run.}
\end{figure}

The energy calibration and resolution of the $\gamma$-ray at the
signal energy is extracted from the two CEX periods.
Figure~\ref{fig:CEX spectrum} shows an example of the energy
spectrum measured with $54.9$~MeV photons from the CEX reaction. A
small correction is made to take into account the different
background present in the LXe volume during the operation with the
pion beam.

The line shape is asymmetric with a low energy tail due to
$\gamma$-rays converting in front of the LXe sensitive volume. A 3D
mapping of the parameters is also made, since they depend to some
extent on the position of the $\gamma$-ray conversion, mainly on the
conversion depth inside the detector ($w$). As an example, the
average resolution for deep events ($w>2$~cm) is measured to be
$\Delta E / E = (5.8\pm0.35)\,\%$~FWHM with a right tail of
$\sigma_{\rm R} = (2.0\pm0.15)\,\%$, where the error quoted includes
the variation over the acceptance. The energy scale is constantly
monitored by looking at the reconstructed $17.67$~MeV energy peak
from CW protons on Li and confirmed by a fit of the photon energy
spectrum to the expected spectra from the $\mu^+ \to e^+ \nu_e
\bar\nu_\mu \gamma$ decay, positron annihilation-in-flight and
$\gamma-$ray pile-up, folded with the line-shape determined during
the $\pi^0$-experiment. The systematic uncertainty on the energy
scale is estimated, by a comparison of these measurements, to be $<
0.4\,\%$.

\subsection{Relative time, $t_{e \gamma}$}
The positron time measured by the scintillation counters is
corrected by the time-of-flight of the positron from the target to
the TC, as measured by the track-length in the spectrometer. The
photon time is determined by the waveforms of the LXe PMTs and
corrected by the line-of-flight that starts from the positron vertex
on the target and ends at the reconstructed conversion point in the
LXe detector.

\begin{figure}
\begin{center}
\includegraphics[width=7cm, bb= 17 15 521 490]{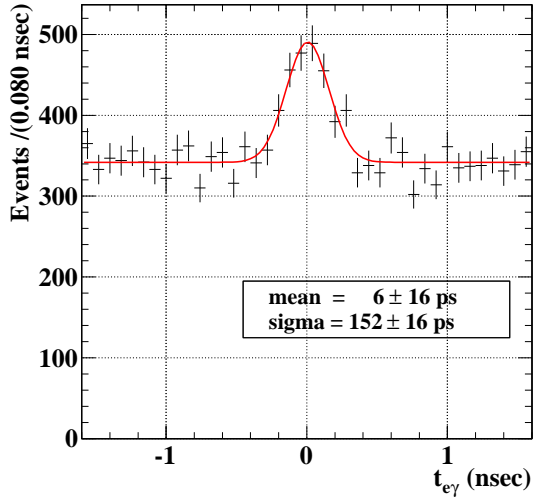}
\end{center}
\caption{\label{fig:dteg}The relative time distribution $t_{e
\gamma}$ showing the RMD peak obtained during physics runs, for $40
< E_\gamma < 45$~MeV. The actual resolution function used in the
analysis takes into account the different gamma energy.}
\end{figure}

In Figure~\ref{fig:dteg}, the relative time distribution between the
positron and the photon in a normal physics run is shown: the RMD
peak (outside of the blinding-box) is clearly visible above the
accidental background. The $t_{e \gamma}$-peak is fitted in the
region of $40 < E_\gamma < 45$~MeV and, by taking into account a
small $E_\gamma$-dependence observed in the $\pi^0$-runs, the timing
resolution for the signal is estimated to be $\sigma_{t_{\rm e
\gamma}} = (148 \pm 17)$~ps. The relative time between the LXe
detector and the TC was monitored over the whole data-taking period
by observing the RMD time peak in runs at normal intensity, and was
found to be stable to within $20$~ps.
\subsection{Relative angles, $\theta_{e \gamma}$ and $\phi_{e \gamma}$}
The positron direction and decay vertex position are determined by
projecting the positron back to the target. The $\gamma$-ray
direction is defined by the line linking its reconstructed
conversion point in the LXe detector with the vertex of the
candidate companion positron. The resolution of the angles between
the two particles is evaluated by combining the angular resolution
and the vertex position resolution in the positron detector and the
position resolution in the photon detector. The positron angular
resolution is evaluated by exploiting tracks that make two turns in
the spectrometer, where each turn is treated as an independent
track. The $\theta$- and $\phi$-resolutions\footnote{taking the
$z$-axis as the beam-axis, $\theta$ is defined as the polar angle,
while $\phi$ is the azimuthal angle} are extracted separately from
the difference of the two track segments at the point of closest
approach to the beam-axis and are $\sigma_\theta = 18$~mrad,
$\sigma_\phi = 10$~mrad. Due to this difference, $\theta_{e \gamma}$
and $\phi_{e \gamma}$ are treated separately in the analysis. The
vertex position resolutions are measured, using the two-turn
technique, to be $\sim 3.2$~mm and $\sim 4.5$~mm in the vertical and
horizontal directions on the target plane respectively. These values
were confirmed independently by a method which reconstructs the
edges of several holes placed in the target.

The position of the photon conversion point is reconstructed by
using the distribution of the light seen by the PMTs near the
incident position. The performance of the position reconstruction is
evaluated by a Monte Carlo simulation and it is validated in a
dedicated CEX experiment by placing a lead collimator in front of
the photon detector. The average position resolutions along the two
orthogonal front-face sides of the LXe detector and the depth
direction ($w$) are estimated to be $\sim 5$~mm and $\sim 6$~mm
respectively.

On combining the individual resolutions, the averaged opening-angle
resolutions of $21$ and $14$~mrad for $\theta_{e \gamma}$ and
$\phi_{e \gamma}$ are obtained respectively.

\section{Data analysis}
The analysis algorithms are calibrated and optimized by means of a
large data sample in the side-bands outside of the blinding-box. The
background level in the signal region can also be studied with the
event distribution in the side-bands since the primary source of
background in this experiment is accidental.

The blinding-box is opened after completing the optimization of the
analysis algorithms and the background study. The number of
$\megsign$ events is determined by means of a maximum likelihood fit
in the analysis window region defined as $46\,{\rm
MeV}<\egamma<60\,{\rm MeV}$, $50\,{\rm MeV}<\epositron<56\,{\rm
MeV}$, $|\tegamma|<1\,{\rm ns}$, $|\theta_{e \gamma}| < 100\,{\rm
mrad}$ and $|\phi_{e \gamma}| <100\,{\rm mrad}$.

\begin{figure}[htp]
\centering
\subfigure{
\includegraphics[width=5.7cm]{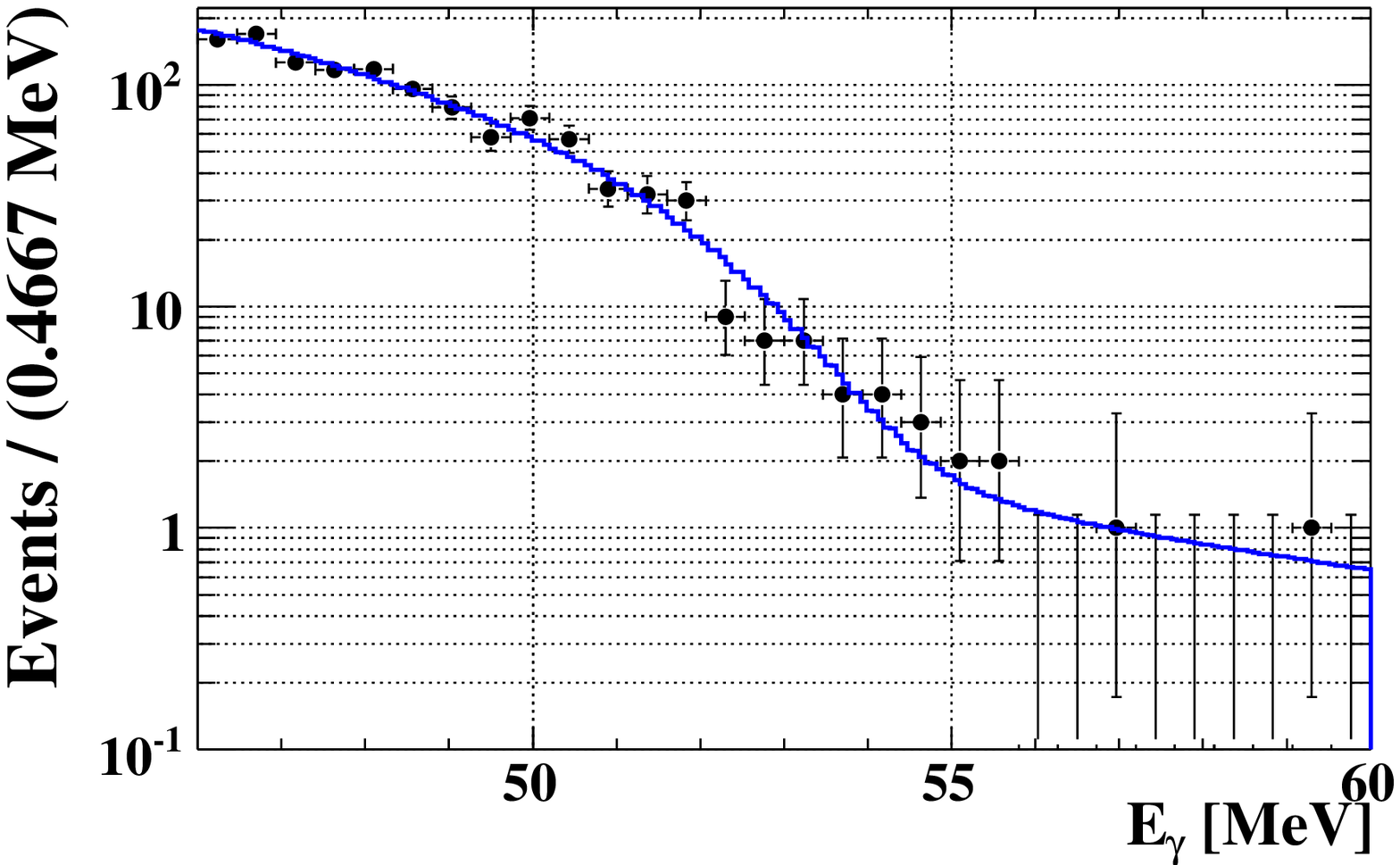}
}
\subfigure{
\includegraphics[width=5.7cm]{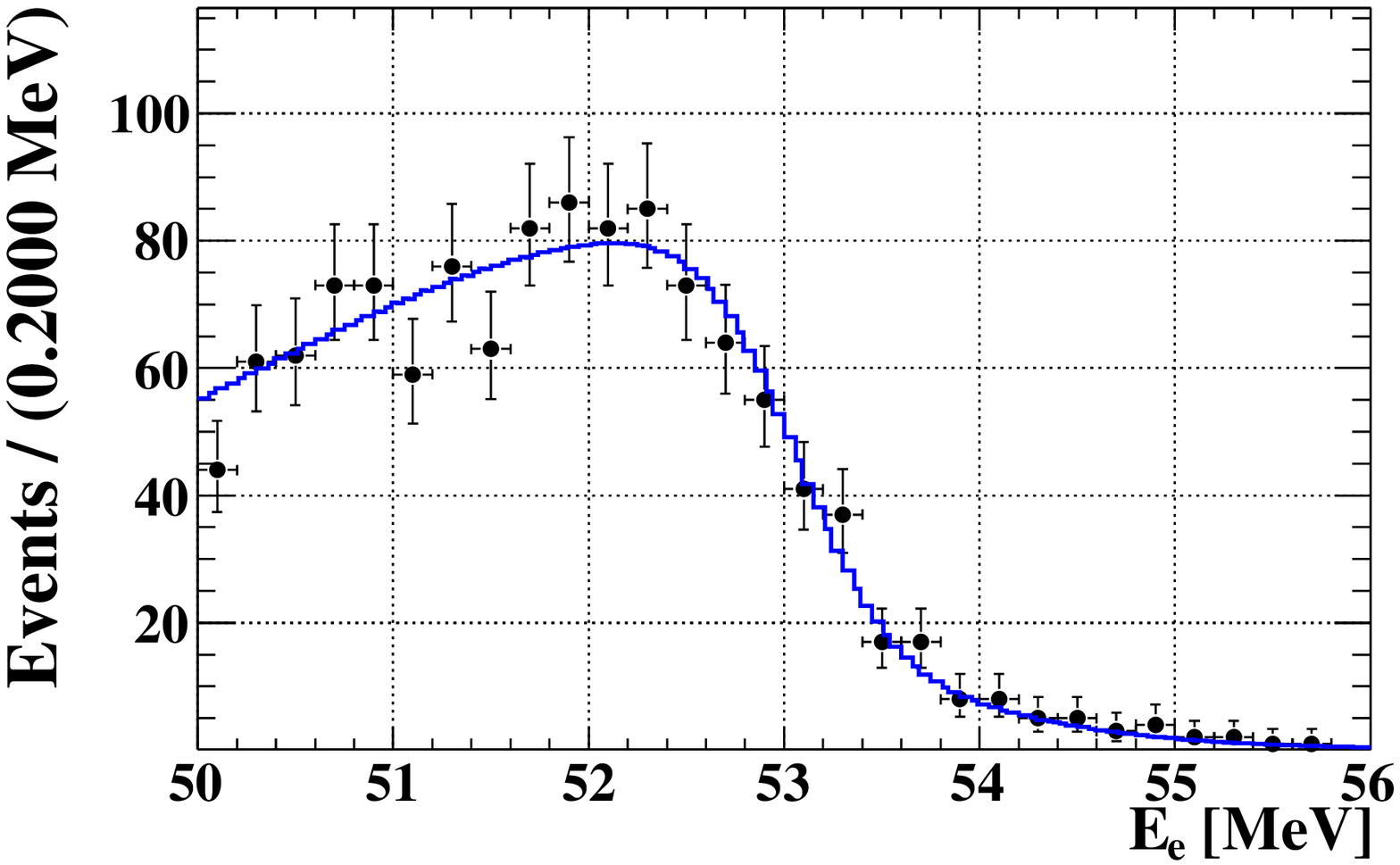}
}
\subfigure{
\includegraphics[width=5.7cm]{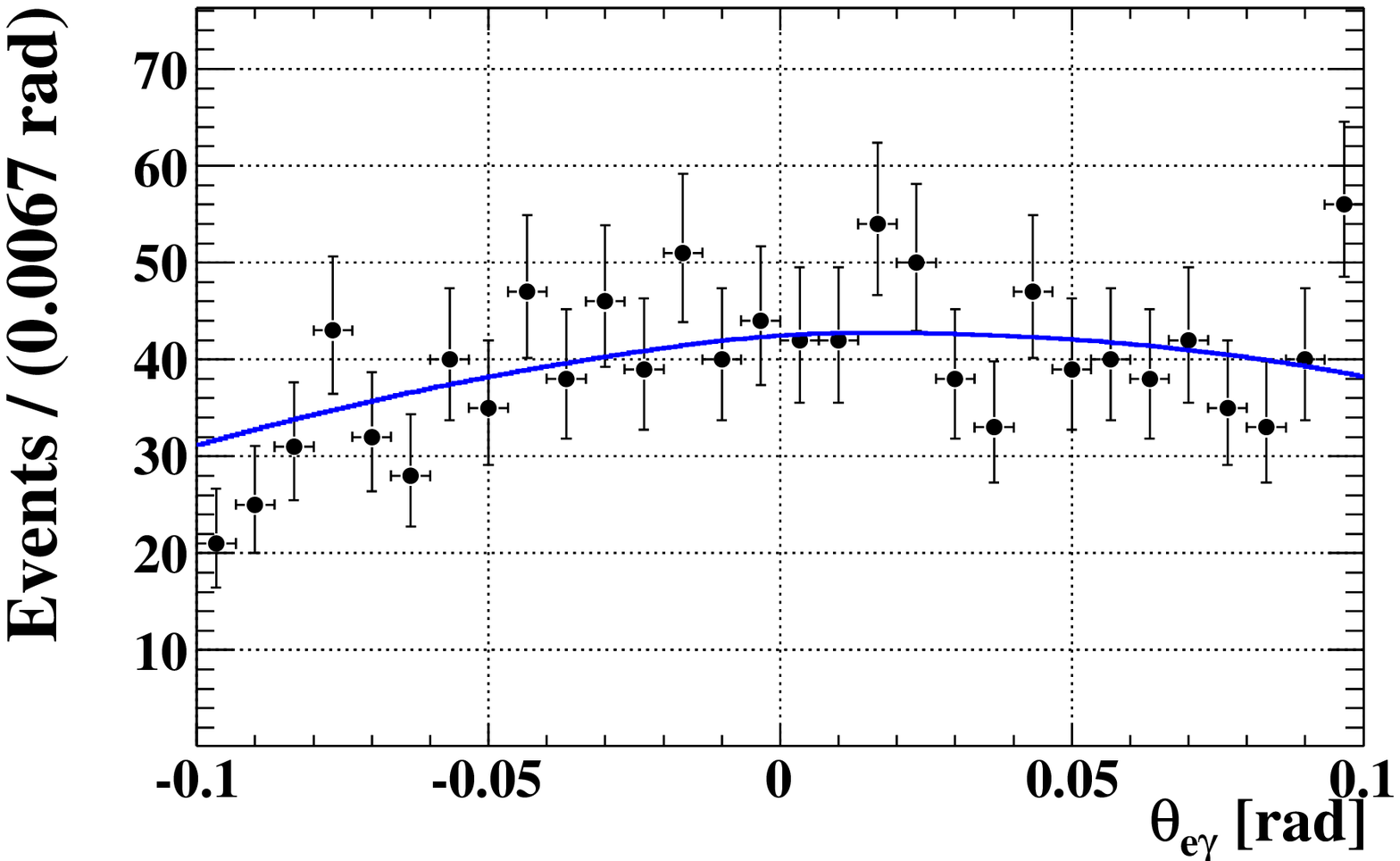}
}
\subfigure{
\includegraphics[width=5.7cm]{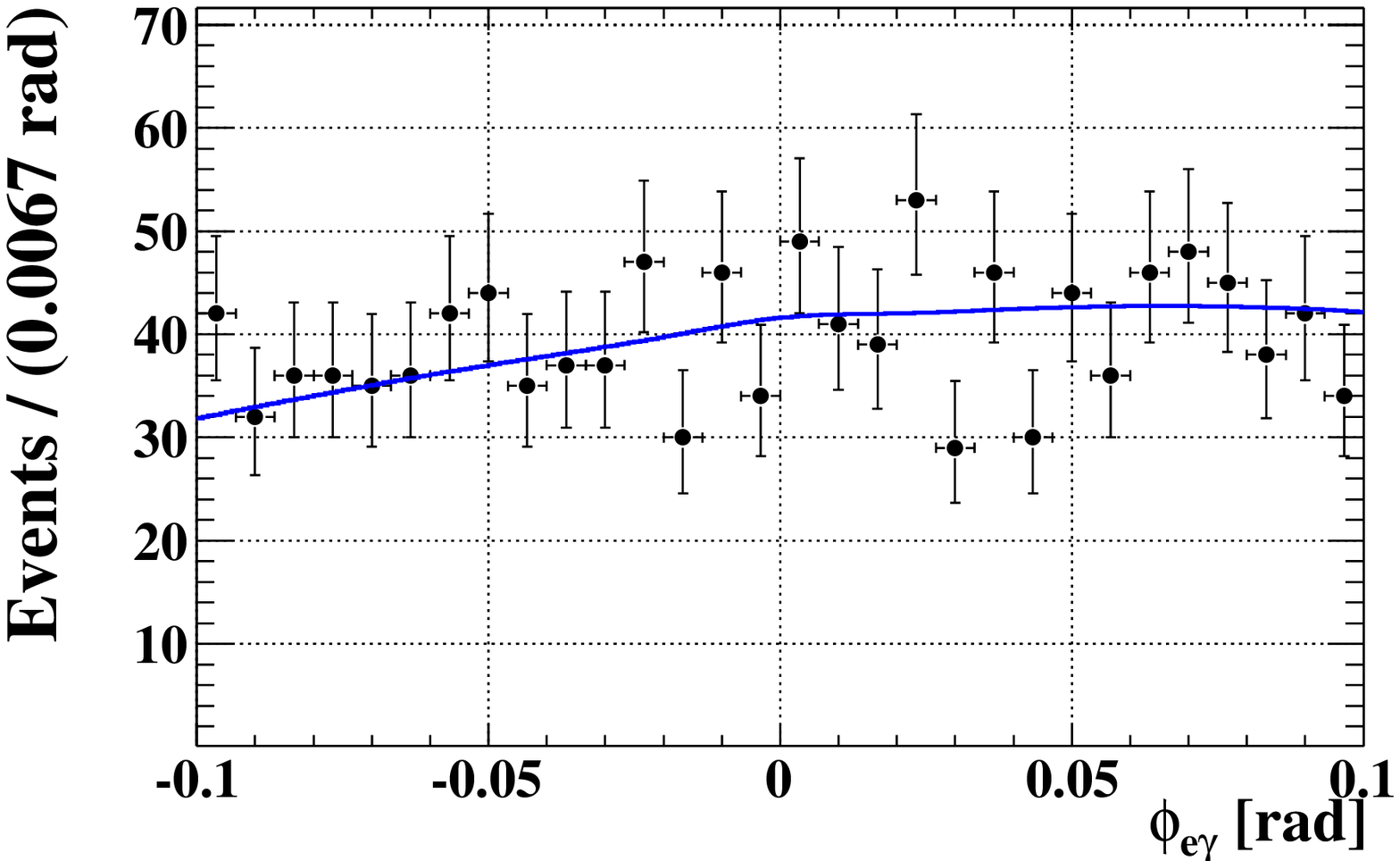}
}
\subfigure{
\includegraphics[width=5.7cm]{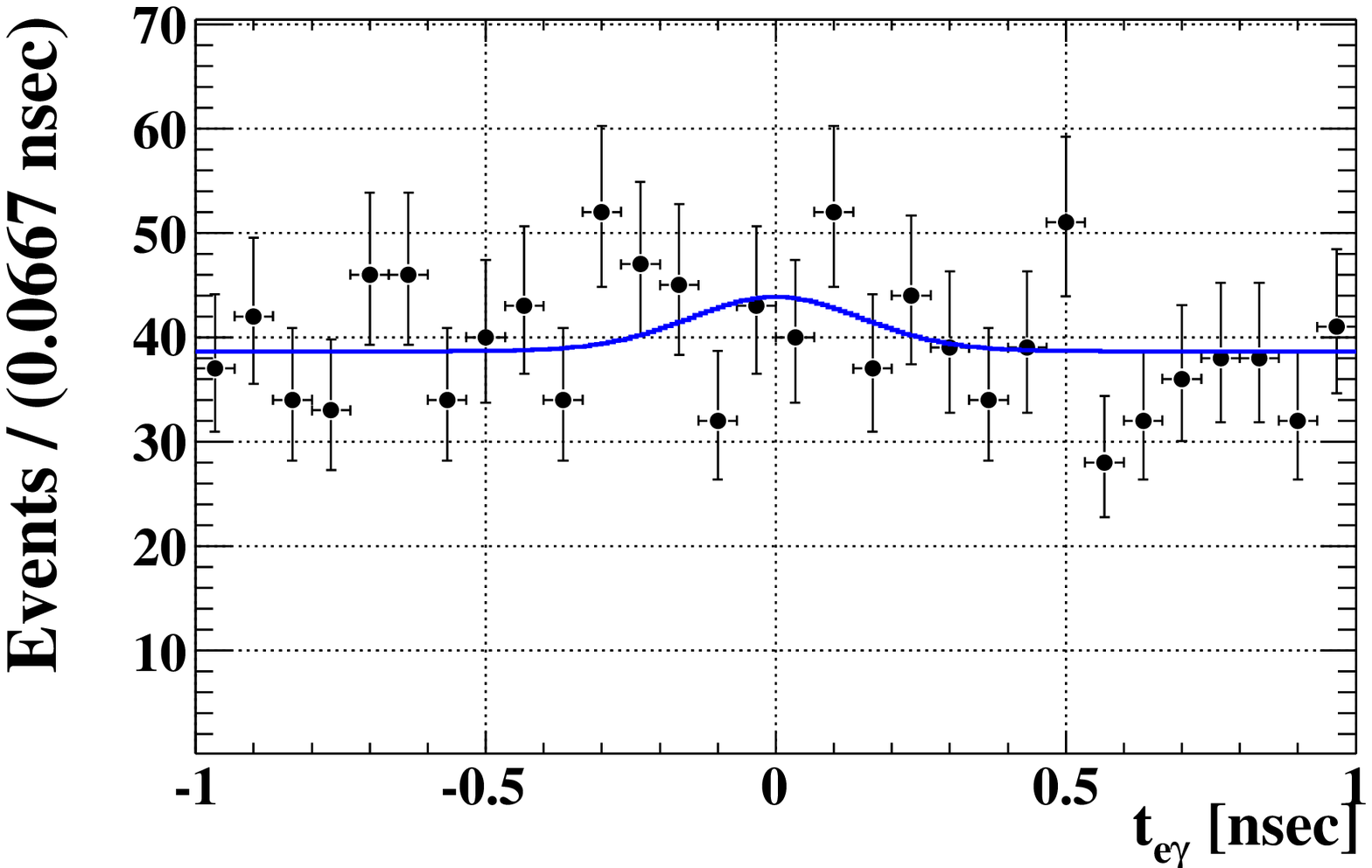}
} \caption{\label{fig:event distribution}Projected distributions for each observable, containing all
events in the analysis window. A solid line shows the likelihood
functions fitted to the data.}
\end{figure}

An extended likelihood function ${\cal L}$ is constructed as,
\begin{eqnarray}
   {\cal L}(\nsig, \nrd, \nbg)
  & = &\frac{N^{\nobs}\exp^{-N}}{\nobs!}\prod_{i = 1}^{\nobs}\left[\frac{\nsig}{N}S+\frac{\nrd}{N}R+\frac{\nbg}{N}B\right], \nonumber
\end{eqnarray}
where $\nsig$, $\nrd$ and $\nbg$ are the number of $\meg$, RMD and
accidental background (BG) events, respectively, while $S$, $R$ and
$B$ are their respective probability density functions (PDFs).
$\nobs (= 1189)$ is defined as the total number of events observed
in the analysis window and  $N = \nsig + \nrd + \nbg$. The signal
PDF $S$ is the product of the statistically independent PDFs for the
five observables ($\egamma$, $\epositron$, $\tegamma$, $\dtheta$ and
$\dphi$), each defined by their corresponding detector response
function with the measured resolutions, as previously described. The
RMD PDF $R$ is the product of the PDF for $\tegamma$, which is the
same as that for the signal and the PDF for the other correlated
observables ($\egamma$, $\epositron$, $\dtheta$ and $\dphi$). The
latter is formed by folding the theoretical RMD spectrum
\cite{kuno-okada} with the detector response functions. The BG PDF
$B$ is the product of the background spectra for the five
observables, which are precisely measured in the data sample in the
side-bands outside the blinding-box. The position dependence of the
resolutions in the case of the $\gamma-$ray is taken into account in
the PDFs, together with all their proper normalizations. The event
distributions of the five observables for all events in the analysis
window are shown in Figure~\ref{fig:event distribution}, together
with the projections of the fitted likelihood function.

The 90\,\% confidence intervals on $\nsig$ and $\nrd$ are determined
by the Feldman-Cousins approach~\cite{Feldman-Counsins}. A contour
of 90\,\% C.L. on the ($\nsig$, $\nrd$)-plane is constructed by
means of a toy Monte Carlo simulation. On each point on the contour,
90\,\% of the simulated experiments give a likelihood ratio ($ {\cal
L}/{\cal L_{\rm max}}$) larger than that of the ratio calculated for
the data. The limit for $\nsig$ is obtained from the projection of
the contour on the $\nsig$-axis. The obtained upper limit at 90\,\%
C.L. is $\nsig<14.7$, where the systematic error is included.
The largest contributions to the systematic error are from the uncertainty of the selection 
of photon pile-up events, the photon energy scale, the response function 
of the positron energy and the positron angular resolution.
The confidence intervals are calculated by three independent
likelihood fitting tools, each with different schemes and
algorithms. The results are all consistent. The expected number of
RMD events in the analysis window is calculated to be $40\pm8$,
obtained by scaling the number of events in the peak of the
$\tegamma$-distribution, obtained with lower energy cuts, using the
probability ratio in the PDFs. This expectation is consistent with
the best estimate in the likelihood fitting of ($25^{+17}_{-16}$).

The upper limit on ${\rm BR}(\megsign)$ is calculated by normalizing
the upper limit on $\nsig$ to the number of Michel positrons counted
simultaneously with the signal and using the same analysis cuts,
assuming ${\rm BR}(\mu \to e \nu \bar \nu) \approx 1$. This
technique has the advantage of being independent of the
instantaneous beam rate and is nearly insensitive to positron
acceptance and efficiency factors associated with the DCH and TC
detectors, as these differ only slightly between the signal and the
normalization samples, due to small momentum dependent effects. The
branching ratio can in fact be written as:
\begin{eqnarray}
   {\rm BR} (\megsign) 
   & = & \frac{\nsig}{\nenunu}
   \times \frac{f^E_{e \nu \bar \nu}}{P}
   \times \frac{\epsilon^{\rm trig}_{e \nu \bar \nu}}{\epsilon^{\rm trig}_{e \gamma}}
   \times \frac{A^{\rm TC}_{e \nu \bar \nu}}{A^{\rm TC}_{e \gamma}}
   \times \frac{\epsilon^{\rm DCH}_{e \nu \bar \nu}}{\epsilon^{\rm DCH}_{e \gamma}}
   \times \frac{1}{A^{\rm g}_{e \gamma}}
   \times \frac{1}{\epsilon_{e \gamma}},
\nonumber
\end{eqnarray}
where $\nenunu = 11414$ is the number of detected Michel positrons
with $50\,{\rm MeV}<\epositron<56\,{\rm MeV}$; $P = 10^{7}$ is the
prescale factor in the trigger used to select Michel positrons;
$f^E_{e \nu \bar \nu} = 0.101 \pm 0.006$ is the fraction of the
Michel positron spectrum above $50$~MeV; ${\epsilon^{\rm trig}_{e
\gamma}}/{\epsilon^{\rm trig}_{e \nu \bar \nu}} = 0.66 \pm 0.03$ is
the ratio of signal-to-Michel trigger efficiencies; ${A^{\rm TC}_{e
\gamma}}/{A^{\rm TC}_{e \nu \bar \nu}} = 1.11 \pm 0.02$ is the ratio
of signal-to-Michel DCH-TC matching efficiency; ${\epsilon^{\rm
DCH}_{e \gamma}}/{\epsilon^{\rm DCH}_{e \nu \bar \nu}} = 1.02 \pm
0.005$ is the ratio of signal-to-Michel DCH reconstruction
efficiency and acceptance; ${A^{\rm g}_{e \gamma}} = 0.98 \pm 0.005$
is the geometrical acceptance for signal photons given an accepted
signal positron; ${\epsilon_{e \gamma}} = 0.63 \pm 0.04$ is the
efficiency of photon reconstruction and selection criteria. The
trigger efficiency ratio is different from one due to the imposition
of stringent angle matching criteria at the trigger level. The main
contributions to the photon inefficiency are from conversions before
the LXe active volume and selection criteria imposed to reject
pile-up and cosmic ray events.

The limit on the branching ratio of the $\megsign$\ decay is
therefore
\begin{eqnarray}
   {\rm BR}(\megsign) & \leq & 2.8 \times 10^{-11}\qquad {\rm (90\% C.L.)} \nonumber
\end{eqnarray}
where the systematic uncertainty on the normalization is taken into account.

The obtained upper limit can be compared with the branching ratio
sensitivity of the experiment with these data statistics. The
sensitivity is defined as the upper limit of the branching ratio,
averaged over an ensemble of experiments, which are simulated by
means of a toy Monte Carlo, assuming a null signal and the same
numbers of accidental background and RMD events as in the data. The
branching ratio sensitivity in this case is estimated to be
$1.3\times10^{-11}$, which is comparable with the current branching
ratio limit set by the MEGA experiment~\cite{MEGA}. Given this
branching ratio sensitivity, the probability to obtain the upper
limit greater than $2.8 \times 10^{-11}$ is $\sim5~\%$ if systematic
uncertainties in the analysis are taken into account.

\section{Conclusion and Prospects}

A search for the lepton flavour violating decay $\megsign$ was
performed with a branching ratio sensitivity of $1.3\times10^{-11}$,
using data taken during the first three months period of the MEG
experiment in 2008. With this sensitivity, which is comparable with
the current branching ratio limit set by the MEGA experiment, a
blind likelihood analysis yields an upper limit on the branching
ratio of BR$({\mu^+ \to e^+ \gamma}) \leq 2.8 \times 10^{-11}$ (90\%
C.L.).

The problem of the reduced performance of the drift chambers, due to
high-voltage trips, has been solved and the chambers
functioned successfully during our 2009 run period. Additional
maintenance to the LXe-detector has also resulted in a near optimal
light-yield. Further improvements to the timing counter fibre
detectors and the digitization electronics are in progress.

\section{Acknowledgments}
We are grateful for the support and co-operation provided by PSI as
the host laboratory and to the technical and engineering staff of
our institutes. This work is supported by DOE DEFG02-91ER40679
(USA), INFN (Italy) and MEXT KAKENHI 16081205 (Japan).
%
%
%

\end{document}